\documentclass[%
 reprint,
 amsmath,amssymb,
 aps,
]{revtex4-1}
\usepackage{braket}
\usepackage{mathrsfs}
\usepackage{caption}
\usepackage{subcaption}
\usepackage{color}
\usepackage{graphicx}
\usepackage{dcolumn}
\usepackage{bm}
\usepackage[dvipsnames]{xcolor}

\begin{document}

\preprint{APS/123-QED}

\title{Floquet topological phase in a generalized $PT$-symmetric lattice}

\author{Elizabeth Noelle Blose}
\affiliation{%
Department of Physics and Astronomy, Eastern Kentucky University Richmond, KY 40475, USA
}%

\date{\today}
\begin{abstract}
We consider a driven, non-Hermitian generalization of the Aubry-Andre-Harper (AAH) model.  We show that the introduction of periodic driving allows us to obtain fully real quasienergy spectra in configurations where the corresponding static system has complex energy eigenvalues for any non-Hermitian degree.   We illustrate how generalized parity-time ($PT$) symmetry can be applied within the Floquet formalism and show that our model's fully real quasienergy spectrum corresponds to an unbroken symmetry.  This model exhibits a Floquet topological phase, few examples of have been found in non-Hermitian potentials.
\end{abstract}

\pacs{Valid PACS appear here}
\maketitle

\section{\label{sec:Introduction}Introduction}

Floquet topological phases are of great current interest.  It has been shown that a wide range of topological phases in lattices with static potentials
possess analogs in systems with periodically driven potentials (``Floquet'' systems); furthermore, some of these Floquet systems exhibit topological states that do not exist in static systems \cite{Floquet_topological_insulator, Floquet_topological_insulator_periodic_table,Floquet_3D_topological_insulator,Floquet_various_types_of_topological_phases,Floquet_topological_characterization,Floquet_Aubry_Andre, Floquet_Majorana_proof,Floquet_Majorana_localized_over_time,Floquet_Majorana_other_example,More_Floquet_Majorana}. Floquet topological phases have also been observed experimentally in photonic systems \cite{Rechtsman2013_includes_experiment,Photonic_topo1,Photonic_topo2,Photonic_topo3}, single-photon quantum walks \cite{Kitagawa2012,Quantum_walk_topo2}, and ultracold atoms \cite{Ultracold_topo1}.

For systems with periodically varying potentials, the long-time behavior is governed by Floquet quasienergies, which are constant over time and analogous to the energy eigenvalues governing the behavior of static systems.  Like the energy eigenvalues of static systems, quasienergies must be real in order to conserve probability.  In order to guarantee the reality of the energy spectrum for static systems, or the reality of the quasienergy spectrum for Floquet systems, it is conventional to require that the Hamiltonian of the system be Hermitian.  However, the requirement of Hermiticity can be relaxed.  It was first shown in static systems that non-Hermitian Hamiltonians can yield real energy spectra if the system is parity-time ($PT$) symmetric, meaning that the Hamiltonian commutes with the combined parity-time reversal operator \cite{Bender_PT_intro}.  $PT$-symmetric systems have two phases: a phase of ``unbroken'' symmetry, where energies are fully real, and a phase of ``broken'' symmetry, where (quasi)energies come in complex-conjugate pairs. $PT$-symmetric systems have been experimentally realized in photonic waveguides \cite{optical1,optical3,optical4,optical5,optical6, Zeuner2015, Pan2018,Eichelkraut2013} and laser systems \cite{laser1,laser2,laser3,laser4}, among others \cite{Microcavities,RLC}.  In static $PT$-symmetric systems, a number of topological phases have also been predicted, some of which have no Hermitian analogs \cite{Yuce, Joglekar,nonH_SSH,Top_Band_Theory,Top_ins,nonH_Berry,Top_Num,PTsymm_breaking,Kawabata, Non_Hermitian_Topo_New, Zeuner2015, Pan2018, NewRequestedRef,NonH_Topo_Yao_Wang}.  The investigation of $PT$-symmetric systems with topological phases has recently broadened to include Floquet systems \cite{Yuce_topological_Floquet, Other_Floquet_PT_symm_topological_phases, NonH_Floquet_Topo_2,NonH_Floquet_Topo_3,NonH_Floquet_Topo_4,NonH_Floquet_Topo_5,NonH_Floquet_Topo_6}. 
 As discussed above, Floquet topological phases have proved a rich area of study \cite{Floquet_topological_insulator, Floquet_topological_insulator_periodic_table,Floquet_3D_topological_insulator,Floquet_various_types_of_topological_phases,Floquet_topological_characterization,Floquet_Aubry_Andre, Floquet_Majorana_proof,Floquet_Majorana_localized_over_time,Floquet_Majorana_other_example,More_Floquet_Majorana,Rechtsman2013_includes_experiment,Photonic_topo1,Photonic_topo2,Photonic_topo3,Kitagawa2012,Quantum_walk_topo2,Ultracold_topo1}; however, most investigations of Floquet topological phases to date have taken place in the context of Hermitian dynamics.

In this work, we extend the definition of $PT$ symmetry from
Ref.\,\cite{Bender_PT_intro} to the Floquet formalism and generalize our definition to a class of operators analogous to $PT$, similar to previous studies of generalized $PT$ symmetry in static systems \cite{Generalized_PT}.   We consider a time-dependent generalization of the Aubry-Andr\'e-Harper (AAH) model \cite{Aubry_Andre, Harper} that possesses an analog of $PT$ symmetry. The off-diagonal AAH model is a 1D tight-binding lattice with spatially periodic (or quasiperiodic) tunneling elements \cite{AAH_Hermitian}.
The Hamiltonian of an $N$-site, off-diagonal AAH lattice is given by 
\begin{align}H_0=\sum_{n=1}^{N-1}t_n a_n^\dagger a_{n+1}+h.c.,\nonumber \\
t_n=-J(1+ \lambda \, \text{cos}(2\pi \beta n+ \Phi)) ,\label{eq:H0}
\end{align}
where $a_n^\dagger$ and $a_n$ are the creation and annihilation operators for a fermion at lattice site $n$. Here, $J$ sets the energy scale of tunneling, while $\lambda$ is the tunneling modulation strength, $\Phi$ gives the tunneling phase, and $\beta$ determines the tunneling period.  Configurations with rational values of $\beta$  have spatially periodic tunneling elements and are therefore termed ``commensurate'' models.  Configurations with irrational values of $\beta$ have quasi-periodic tunneling elements and are called ``incommensurate'' models.  This model exhibits a rich array of physical phenomena.  For rational $\beta$ of the form $\beta=1/p$, where $p$ is an integer, the energy spectrum has $p$ bands.  When $p$ is even, zero energy modes exist for ranges of $\Phi$ determined by the values of $p$ and $N$, and furthermore these energy modes correspond to topologically protected states located to the edges of the lattice \cite{AAH_Hermitian, Yuce}. (For example, Ref.\,\cite{Yuce} shows that for $\beta=1/2$, $0 \leq \Phi \leq 2 \pi$ gives topological modes when $N$ is odd, and  $\frac{\pi}{2} \leq \Phi \leq \frac{3 \pi}{2} $ gives topological modes when $N$ is even.)  For irrational $\beta$, the spectrum has a fractional number of bands, and states within the band gaps are localized to the edges.  

Refs.\,\cite{Yuce, Joglekar} discuss a non-Hermitian generalization of the AAH model.  Their lattice has the same tunneling profile as the original off-diagonal AAH lattice, but two imaginary-energy defects are placed at reflection-symmetric sites.  For this system, the reality of the energy spectrum is preserved for particular defect locations and lattice sizes, and the topological modes of the commensurate model persist in the presence of imaginary-energy defects.  

In this work, we demonstrate that by periodically driving the strengths of the non-Hermitian impurities, we obtain real quasienergy spectra for configurations whose static analogs have complex spectra for any strength of non-Hermitian defect.  In section II, we introduced the model and give a brief overview of the Floquet formalism.  In section III, we show how the reality of the quasienergy spectrum can be explained in terms of a symmetry of the Hamiltonian.  To do this, we show how the definition of $PT$ symmetry can be applied within the Floquet formalism and that a fully real quasienergy spectrum corresponds to an unbroken symmetry analogous to $PT$ symmetry.  In section IV, we show that the mid-gap quasienergy modes of the system are topologically protected.  As in the static AAH lattice, these topological modes are closely connected to Majorana bound states \cite{AAH_Hermitian, Kitaev_2001}.

\section{\label{sec:Model and the Floquet formalism}Model and the Floquet formalism}

We consider a time-dependent generalized AAH lattice whose Hamiltonian is of the form
\begin{gather}
 H(t)=H_0+V(t), \nonumber \\ 
V(t)=i \gamma \, \text{cos}(\omega t) (a_{m_0}^\dagger a_{m_0} - a^\dagger_{\bar{m}_0}a_{\bar{m}_0}). \label{eq:H}
\end{gather}
where $H_0$ is the static AAH Hamiltonian given in Eq.\,\ref{eq:H0} and $V(t)$ is a time-dependent perturbation giving gain and loss.

Like the unperturbed AAH lattice, tunneling elements are spatially periodic (quasi-periodic) for rational (irrational) $\beta$. All onsite energies are zero, except at the location of two balanced gain and loss impurities whose onsite energies are periodically modulated in time.  As in the static analog of this lattice discussed in \cite{Yuce, Joglekar}, these impurities are placed symmetrically in the lattice, i.e. for a system with $N$ lattice sites, the impurities are placed at sites ${m_0}$ and $\bar{m_0}=N-{m_0}+1$.   The static model discussed in Ref.\,\cite{Yuce, Joglekar} can be recovered by setting $\omega=0$.  The imaginary-energy defects represent gain and loss sites when the magnitudes of their coefficients are positive and negative, respectively.  
The presence of these defects causes the Hamiltonian to be non-Hermitian.  

Equation \ref{eq:H} could be implemented experimentally structures of coupled optical waveguides.  In these systems, the tunneling strengths can be tuned by varying the spacing of the waveguides (as in \cite{Zeuner2015,optical3}, for example), which would allow for the implementation of the static component of the Hamiltonian $H_0$.  Gain and loss have been widely implemented in photonic waveguide systems by using, for example, complex indices of refraction \cite{ optical4,optical5,Zeuner2015, optical6, Pan2018,optical3, optical1} or by modulating the shape of certain waveguides \cite{Pan2018, Eichelkraut2013}. Spatially modulating the gain and loss would realize the time-varying part of the Hamiltonian $V(t)$.  In some implementations, it may even be unnecessary to include gain; adding a background loss term to Equation \ref{eq:H} would create a system analogous to the passive $PT$ lattices implemented in \cite{Pan2018,optical1,optical6, optical5, Eichelkraut2013}.  In our original model, a fully real quasienergy spectrum corresponds to a phase of unbroken symmetry (as will be discussed in Section \ref{sec:operator}). Once a background of loss is added, the phase of unbroken symmetry would correspond to a slower rate of decay than the phase of broken symmetry \cite{optical1}.  

The behavior of a periodically driven system can be characterized by the system's Floquet quasienergies, which are analogous to the energy eigenvalues of a static potential, and Floquet states, which describe the evolution of the system over time \cite{Shirley, Sambe, Floquet_more}.   In order to determine the Floquet quasienergies (and, in Section \ref{sec:micromotion}, the Floquet states) of the system, we employ the methods of Refs. \cite{Shirley, Sambe, Floquet_more}.  The Floquet theorem states that, given a periodic potential $V(t)$ of period $T=2\pi/\omega$, the solutions of the time-dependent Schr\"{o}dinger equation
\begin{equation}
i \frac{d}{dt}\ket{\psi(t)} = H(t) \ket{\psi(t)} 
\end{equation}
are given by $\ket{\psi_\alpha(t)}=e^{-i \epsilon_\alpha t} \ket{\phi_\alpha(t)}$ with $\ket{\phi_\alpha(t)}= \ket{\phi_\alpha(t+T)}$. (We work in units where $\hbar=1$.) $\epsilon_\alpha$ and $\ket{\phi_\alpha(t)}$ are termed the Floquet quasienergies and Floquet modes, respectively.  As the form of this solution suggests, quasienergies must be real in order to conserve probability.  
To solve this equation, we define the Floquet Hamiltonian 
\begin{equation}
H_F=H(t)-i \frac{d}{dt},\label{eq:H_F}
\end{equation}
which satisfies 
\begin{equation}
H_F \ket{\phi_\alpha(t)}=\epsilon_\alpha \ket{\phi_\alpha(t)}. \label{eq:fake_eval}
\end{equation} 
The Floquet Hamiltonian can be rewritten in the frequency domain as 
\begin{align}
\mathcal{H}^{q,r}_{n,m}= -J(1+ \lambda\, \text{cos}(2\pi \beta n+ \Phi)) \delta_{q,r} (\delta_{n,m+1}-\delta_{n,m-1}) \nonumber \\ -i q \omega \delta_{n,m}\delta_{q,r} +i \frac{\gamma}{2} \delta_{m,n}(\delta_{n,m_0}-\delta_{n,\bar{m}_0})(\delta_{q,r+1}+\delta_{q,r-1}),  \label{eq:Fourier}
\end{align} 
where $m$, $n$ are site indices and $q$, $r$ are Floquet band indices, sometimes called ``photon sectors.'' These photon sectors arise because $\epsilon_\alpha$ and $\ket{\phi_\alpha(t)}$ are not uniquely defined: if $\epsilon_\alpha$ is a Floquet quasienergy, then so is $\epsilon_{\alpha,q}=\epsilon_\alpha+q \omega$ for any integer $q$.  The Floquet mode corresponding to $\epsilon_{\alpha,q}$ is $\ket{\phi_{\alpha,q}(t)}=e^{i q t\omega}\ket{\phi_{\alpha}(t)}$.  Therefore, the eigenvalues of Eq.\,\ref{eq:Fourier} can be labeled $\epsilon_{\alpha,q}$ to account for the ambiguity in their definition. 
The eigenvectors are the Floquet modes translated to the extended Hilbert space $\mathcal{F}=\mathcal{H} \otimes \mathcal{L}_T$, where $\mathcal{H}$ is the state space of the system and $\mathcal{L}_T$ is the space of square-integrable periodic functions with period $T=2\pi/\omega $\cite{ Shirley, Sambe, Floquet_more}. In $\mathcal{F}$, the Floquet modes are written as $|\phi_{\alpha,q}\rangle\rangle$ and have an inner-product defined by $\langle \langle \phi_{\alpha,q}|\phi_{\beta,r}\rangle\rangle=\frac{1}{T}\int_0^T \braket{\phi_{\alpha,q}| \phi_{\beta,r}} dt =\delta_{\alpha,\beta}\delta_{q,r}$.   
When the quasienergies are real, the magnitude of the full time-averaged Floquet state (written in $\mathcal{F}$-space) $|\langle \langle \psi_{\alpha,q}|\psi_{\alpha,r}\rangle\rangle|^2$ follows easily from these eigenvectors, since $|\langle \langle \psi_{\alpha,q}|\psi_{\alpha,q}\rangle\rangle| ^2=| \langle \langle \phi_{\alpha,q}|\phi_{\alpha,q}\rangle\rangle|^2$ for real $\epsilon_{\alpha,q}$.  Therefore, by diagonalizing $\mathcal{H}$, we can obtain the Floquet quasienergies and, when the quasienergies are real, the magnitudes of the time-averaged Floquet states.

\begin{figure*}[t!]
    \centering
    \begin{subfigure}{0.3\textwidth}
        \centering
        \caption{}
        \includegraphics[scale=0.28]{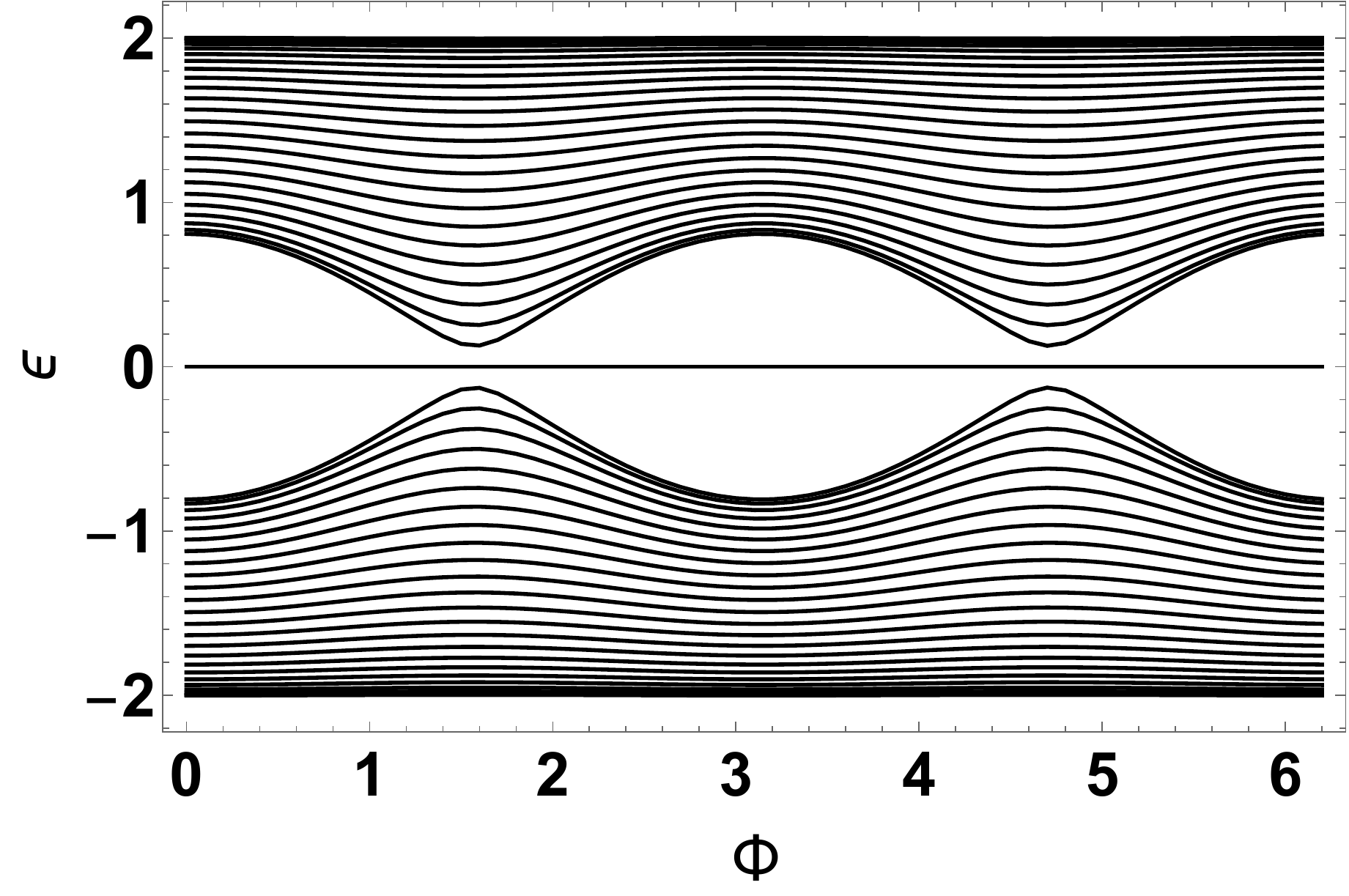}

    \end{subfigure}%
    ~ 
    \hspace{.1cm}
    \begin{subfigure}{0.3\textwidth}
        \centering
        \caption{}
        \includegraphics[scale=0.28]{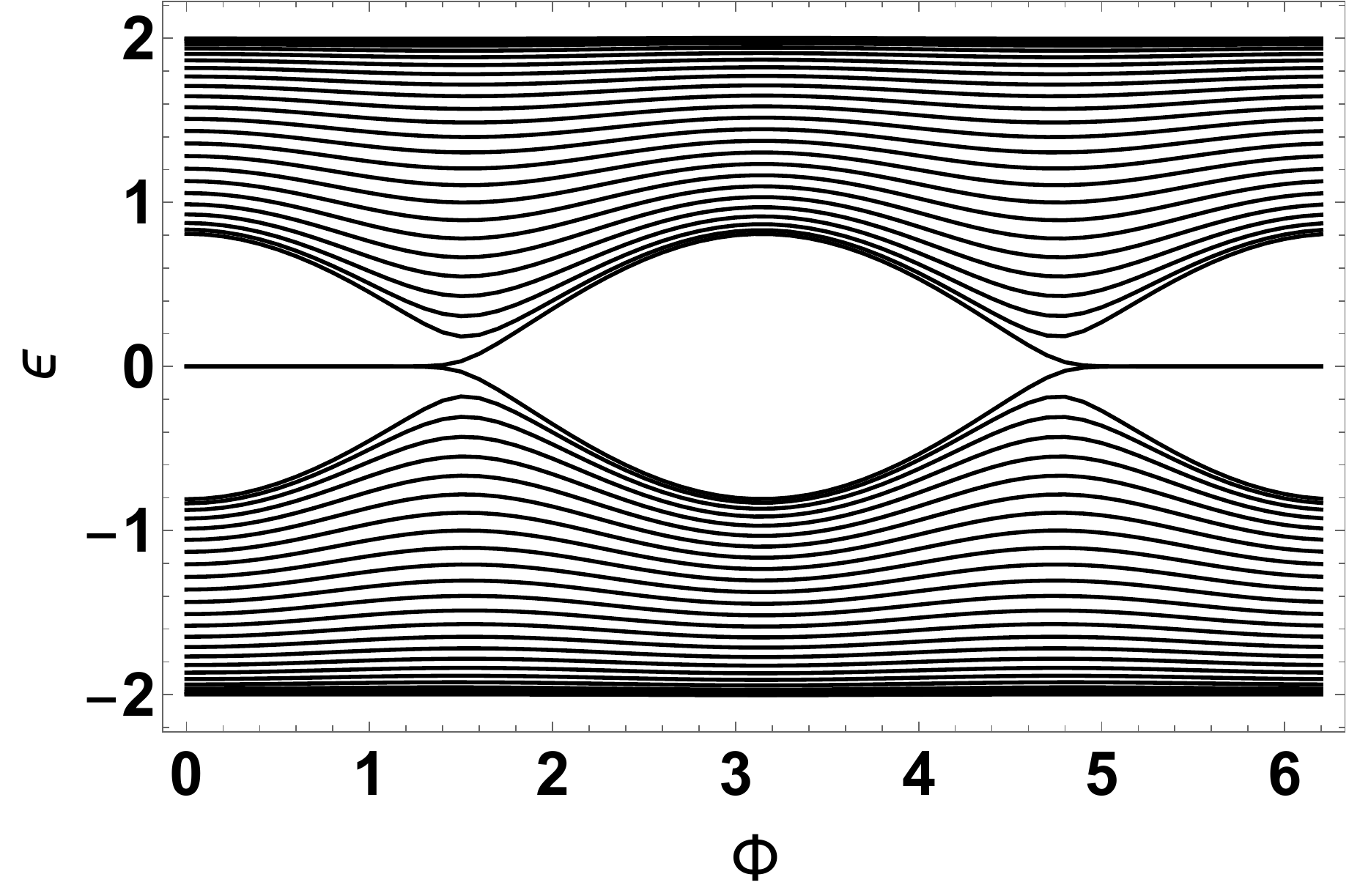}
    \end{subfigure}
       ~ 
    \hspace{.1cm}
    \begin{subfigure}{0.3\textwidth}
        \centering
        \caption{}
        \includegraphics[scale=0.43]{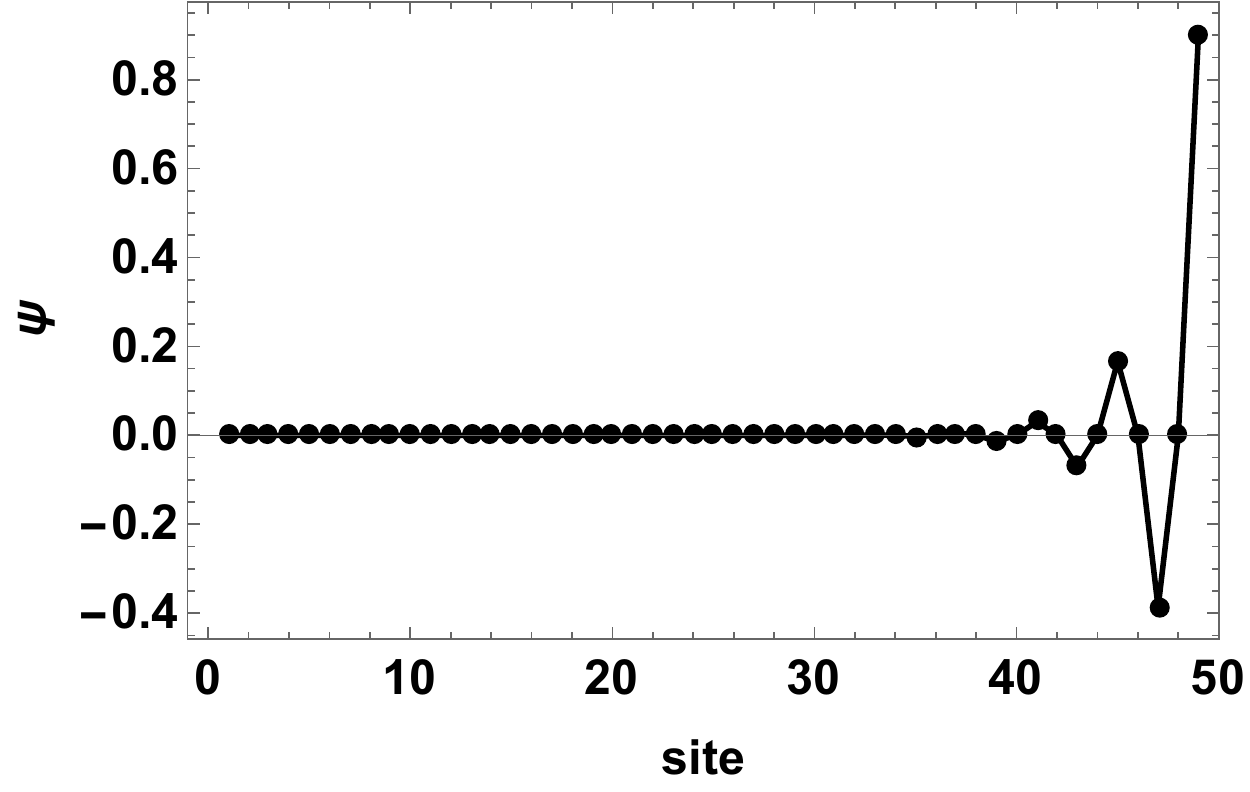}
    \end{subfigure}
    \caption{Quasienergy spectra for $\beta=1/2$, $\lambda=0.4$, $\gamma=3$,   $m_0=3$,  $\omega=2\pi$, and (a) $N=49$, (b) $N=50$.  (c) shows the time-averaged, first Floquet-Brillouin zone mid-gap ($\epsilon=0$) mode corresponding to the spectrum in (a) with $\Phi=0$.  This mode is localized to the first photon sector, and so only that sector is shown. }\label{fig:beta1/2}
\end{figure*}

Because the Floquet quasienergies are periodic modulo $\omega$, all Floquet quasienergies and states can be obtained by considering only the first Floquet-Brillouin zone (otherwise known as the first photon sector), defined as the set of the $N$ lowest-magnitude quasienergies.  
To numerically compute the Floquet quasienergies and time-averaged states, a cutoff index $N_f$ is chosen so that $p,q\leq N_f$ and $\mathcal{H}^{p,q}_{n,m}$ is a $N (2N_f+1)\times N (2N_f+1)$-dimensional matrix.  Numerical errors arise in the highest-magnitude FBZs, and so $N_f$ must be large enough to avoid introducing errors into the first FBZ.   
For the parameters discussed in this work, $N_f=20$ is sufficient to approximate the first Floquet-Brillouin zone eigenvalues and eigenvectors of an infinite-dimensional matrix.  
(In other words, the first FBZ quasienergies and modes do not change as $N_F$ is increased further.) Again, the presence of imaginary-energy defects causes the Hamiltonian to be non-Hermitian (as is evident from the matrix representation of Eq.\,\ref{eq:Fourier}).  
Furthermore, because of the periodicity of the tunneling elements, $H_0$ is not in general parity-symmetric ($P H_0 P^{-1} \neq H_0$), and so the system as a whole does not possess possess $PT$ symmetry ($PT H(t) (PT)^{-1}\neq H$). Nevertheless, this system does have a fully real quasienergy spectrum as long as $\gamma$ is below some threshold.  
An example of a configuration with a real quasienergy spectrum is shown in Fig.\,\ref{fig:beta1/2}.  
Because the lattice does not possess $PT$ symmetry or a Hermitian Hamiltonian, we show in Section III that another physical condition ensures the reality of the quasienergy spectrum. 
%
%
%
%
\section{\label{sec:reality}Reality of the spectrum}

\subsection{\label{sec:operator}Generalized $PT$ symmetry}

Because the Hamiltonian defined in Eq.\,\ref{eq:H} is neither Hermitian nor $PT$-symmetric, another condition must guarantee the reality of the quasienergy spectrum.  
We consider a symmetry of the Hamiltonian  given by an operator $\mathcal{A}$ and show that this symmetry guarantees that quasienergies are either real or come in complex-conjugate pairs, just like $PT$ symmetry in other systems.  
Ref.\,\cite{Bender_PT_intro} shows that, in the context of static systems, $PT$ symmetry guarantees that energy eigenvalues are either real or come in complex-conjugate pairs.  But, this proof holds for any operator $\mathcal{A}$ obeying three conditions: (1) $\mathcal{A}$ commutes with the system's Hamiltonian, (2) $\mathcal{A}$ is antilinear, and (3) the eigenvalues of $\mathcal{A}$ have unit modulus.  Antiunitary operators obey conditions (2) and (3) by definition.  Therefore, for static systems, any antiunitary operator that commutes with the Hamiltonian guarantees that the energies are either real or come in complex-conjugate pairs, and so $PT$ symmetry is simply a special case of a broader phenomenon.  This framework can be extended to the Floquet formalism by requiring that $\mathcal{A}$ commute with $H_F$ rather than the original Hamiltonian, $[H_F,\mathcal{A}]=0$, since Eq.\,\ref{eq:fake_eval} takes the place of the characteristic equation for static systems.  Likewise, condition (3) is replaced by the requirement that the instantanteous eigenvalues of $\mathcal{A}$ have unit modulus.  For the system defined in Eq.\,\ref{eq:H}, $\mathcal{A}$ can be defined as 
\begin{align}
    \mathcal{A} H(t) \mathcal{A}^{-1} =H(t), \nonumber \\
    \mathcal{A} = \mathcal{S} \mathcal{D} T \label{eq:Adef}
\end{align}
where $\mathcal{S}$ is a linear operator representing sublattice symmetry 
\begin{align}
\mathcal{S} a_n \mathcal{S}^{-1}= (-1)^n a_n, \label{eq:Sdef}
\end{align}
$\mathcal{D}$ is a linear operator representing charge-conjugation 
\begin{align}
    \mathcal{D} a_n \mathcal{D}^{-1}= a_n^\dagger, \label{eq:Ddef}
\end{align} 
and $T$ is an antilinear operator representing time-reversal
\begin{align}
    T i T^-1=-i, \nonumber \\
    T (\frac{d}{dt}) T^{-1}=-(\frac{d}{dt}) T.  \label{eq:Tdef}
\end{align}  Therefore,
\begin{align}
    \mathcal{S} H(t) \mathcal{S}^{-1}= -H(t)^* \nonumber \\
    \mathcal{D} H(t) \mathcal{D}^{-1}= -H(t) \nonumber \\
    T H(t) T^{-1}= H(t)^*. \label{eq:SDTaction}
\end{align}
Therefore, $\mathcal{A}$ is antiunitary and has $[H_F,\mathcal{A}]=0$ as required.  

We now extend the proof given in Ref.\,\cite{Bender_PT_intro} to the Floquet formalism.  We first assume that the instantaneous eigenstates of $H_F$ ($\ket{\phi_\alpha (t)}$) are also instantaneous eigenstates of $\mathcal{A}$.  By condition (3), the instantaneous eigenvalue equation at each time $t$,
\begin{equation}\mathcal{A}\ket{\phi_\alpha (t)}=\lambda(t)\ket{\phi_\alpha (t)}, \label{eq:Aeval}
\end{equation}
is solved by an eigenvalue of the form $\lambda(t)= e^{i \theta(t)}$for some real $\theta(t)$.  
Applying $H_F$ to both sides of this equation, we have
\begin{align}
    H_F \mathcal{A} \ket{\phi_\alpha (t)} &=H_F e^{i \theta(t)}\ket{\phi_\alpha (t)}\nonumber\\
    &=e^{i\theta(t)}H_F\ket{\phi_\alpha (t)} \nonumber\\
     &=e^{i\theta(t)}\epsilon_\alpha\ket{\phi_\alpha (t)}\label{eq:eps}
\end{align}
But, $[H_F,\mathcal{A}]=0$, and so
\begin{align}
    H_F \mathcal{A}\ket{\phi_\alpha (t)} &= \mathcal{A}  H_F \ket{\phi_\alpha (t)} \nonumber \\
    &=\mathcal{A} \epsilon \ket{\phi_\alpha (t)} \nonumber\\
    &=\epsilon^* \mathcal{A} \ket{\phi_\alpha (t)} \nonumber\\
    &=\epsilon^* e^{i\theta(t)}\ket{\phi_\alpha (t)}. \label{eq:eps_star}
\end{align}
Combining Eqs.\,\ref{eq:eps} and \ref{eq:eps_star}, we see that $\epsilon=\epsilon^*$.  However, we assumed that the instantaneous eigenstates of $H_F$ were also instantaneous eigenstates of $\mathcal{A}$, which is not necessarily guaranteed by the commutation of $H_F$ and $\mathcal{A}$, since $\mathcal{A}$ is antilinear.  Without this assumption,  Eqs.\,\ref{eq:Aeval}-\ref{eq:eps_star} do not hold, and so the quasienergies are not guaranteed to be real.  In this case, for each quasienergy and Floquet mode pair $\epsilon$ and $\ket{\phi_\alpha(t)}$, there exists a corresponding $\mathcal{A}$-conjugated pair  $\epsilon^*$ and $\mathcal{A}\ket{\phi_\alpha(t)}$.  

 Therefore, as with $PT$ symmetry, we say that $\mathcal{A}$ symmetry is unbroken when the instantaneous eigenstates of the Floquet Hamiltonian are simultaneous eigenstates of $\mathcal{A}$.  As shown above, unbroken $\mathcal{A}$ symmetry guarantees that the quasienergies are  real.  When $\mathcal{A}$ symmetry is broken, the Floquet modes are no longer instantaneous eigenstates of $\mathcal{A}$, and quasienergies occur in complex-conjugate pairs.   
 
For the Hamiltonian defined in Eq.\,\ref{eq:H}, $\mathcal{A}$ is as defined in Eq.\,\ref{eq:Adef} and $\mathcal{A}\neq PT$.  But, $\mathcal{A}$ also obeys the definition of a generalized $PT$ symmetry given in Ref.\,\cite{Generalized_PT} for static systems, once again with $H_F$ taking the place of the ordinary static Hamiltonian and the Floquet modes taking the place of the Hamiltonian's eigenstates.  Therefore, we say that this system possesses a generalized $PT$ symmetry.  It should also be noted that $\mathcal{A}$ is a symmetry of the static version of this lattice $H_0$.  So, the reality of the static lattice's energy spectra (Refs.\,\cite{Yuce, Joglekar}) can also be described in terms of the preservation of $\mathcal{A}$ symmetry.  

Given that $\mathcal{A}$ is the symmetry of the static analog of our system (which is still non-Hermitian), one might also wonder whether $A$ could be a symmetry of a non-Hermitian \textit{diagonal} AAH model.  Ref.\,\cite{Yuce} proposed a non-Hermitian generalization of the AAH model given by
\begin{align}
    H_{\text{diag}}= -t\sum_{n=1}^{N-1} a_n^\dagger a_{n+1}+h.c. \nonumber  + \sum_{n=1}^N V \text{cos}(2\pi \beta n+ \Phi) a_n^\dagger a_n  \nonumber \\ + i \gamma (a_{m_0}^\dagger a_{m_0} -a^\dagger_{\bar{m}_0}a_{\bar{m}_0}).
\end{align}
This system does not possess $\mathcal{A}$ symmetry because the addition of the real onsite energy does not conserve $\mathcal{S}$ symmetry.

\begin{figure*}[t!]
    \centering
    \begin{subfigure}{0.5\textwidth}
        \centering
        \caption{}
        \hspace{-1.5cm}
        \includegraphics[scale=0.53]{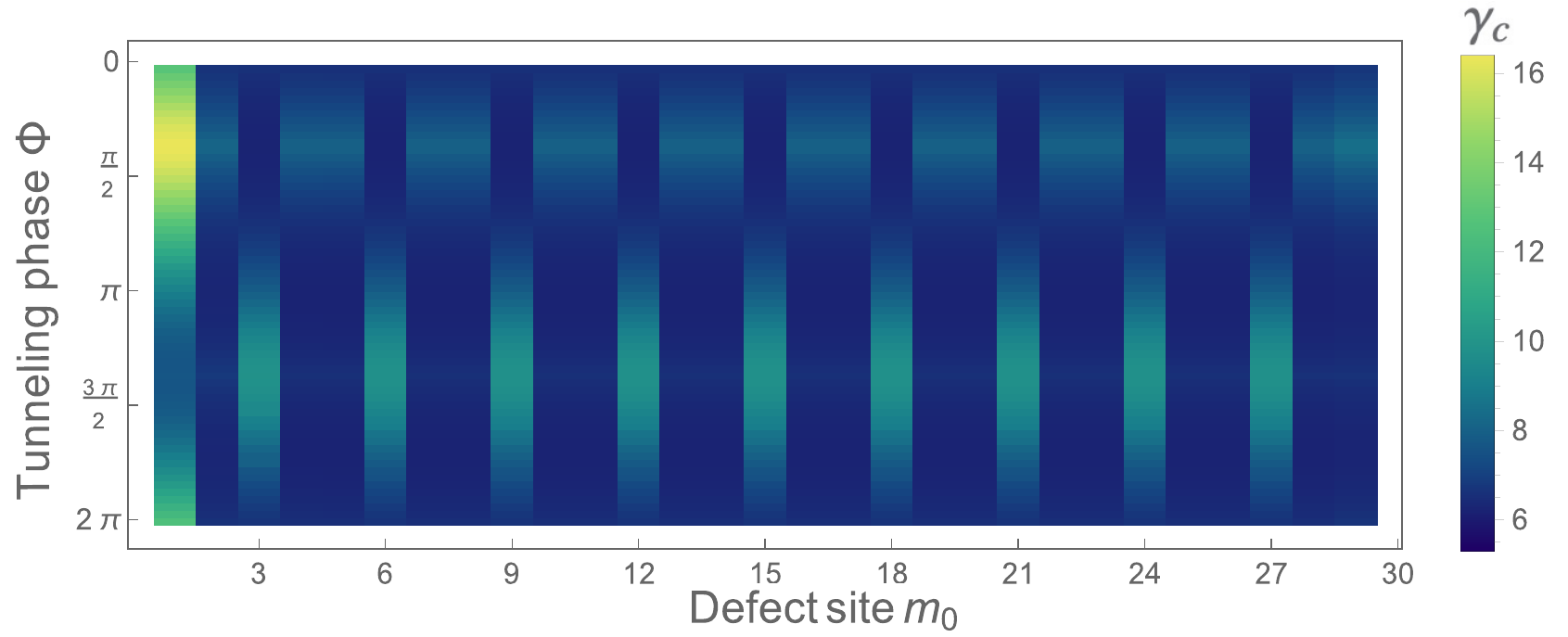}
        \hspace{-2cm}
    \end{subfigure}%
    ~ 
    \hspace{.1cm}
    \begin{subfigure}{0.5\textwidth}
        \centering
        \caption{}
        \includegraphics[scale=0.53]{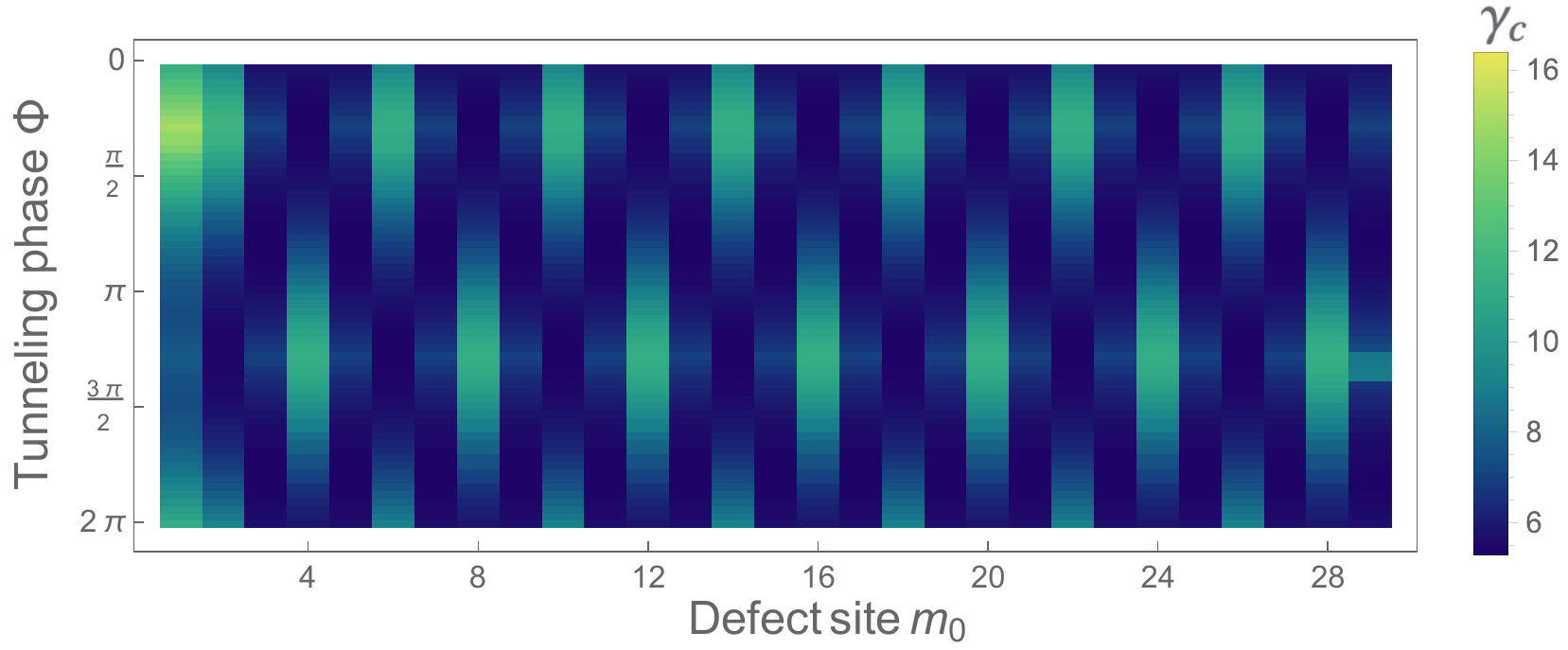}
    \end{subfigure}
    \caption{$\mathcal{A}$ symmetry-breaking threshold $\gamma_c$ as a function of gain site location $m_0$ and tunneling amplitude phase $\Phi$.  Here,  $N=59$, $\omega=2\pi$, $\lambda=0.5$ and $\beta$ is a rational number of the form $q/p$.  (a) has $\beta=1/3$ and (b) has $\beta=1/4$.  In both cases, $\gamma_c$ is approximately periodic with period $p$, except for defect placement near the edges or the center of the lattice.} \label{fig:criticalvalues}
\end{figure*} 
\subsection{Numerical results}
We numerically compute the Floquet quasienergies of the generalized off-diagonal AAH system defined in Eq.\,\ref{eq:H} to obtain a system with a fully real Floquet quasienergy spectrum, which corresponds to the preservation of $\mathcal{A}$ symmetry, and Floquet topological phase. 
We first consider models with rational $\beta$ of the form $q/p$, so that hopping elements are spatially periodic.  Analogous to the static version of the lattice $H_0$ (discussed in Refs. \cite{Yuce, Joglekar}), the spectrum has $p$ bands within each Floquet Brillouin zone.  When mid-gap modes are present, they are localized to the edges of the lattice.   Fig.\,\ref{fig:beta1/2} provides an example quasienergy spectrum and mid-gap mode for $\beta=1/2$.  This system has fully real quasienergies, while the static analog has complex energy eigenvalues for any nonzero $\gamma$.

In the static case, the reality of the spectrum depends strongly on the number of lattice sites and on the placement of the defects: a real energy spectrum may only be obtained when defect location ${m_0}$ obeys ${m_0}=0$ mod $p$ and lattice size $N$ obeys $N+1=0$ mod $p$.  When these conditions are not met, $\mathcal{A}$ symmetry is broken and the spectrum is complex for all $\gamma$.  If $\gamma$ is decreased, the imaginary parts of the spectrum decrease proportionally, but the spectrum is never fully real for nonzero $\gamma$.  When these conditions are met, the energy spectrum is fully real for $\gamma$ below some critical value $\gamma_c$, and complex quasienergies (in complex-conjugate pairs) appear for $\gamma>\gamma_c$. Thus, $\gamma_c$ represents a phase transition between $\mathcal{A}$-broken and $\mathcal{A}$-unbroken symmetry regions. The value of $\gamma_c$ also depends on $m_0$, $\Phi$, $\beta$ and $\lambda$.  Ref.\,\cite{Joglekar} explains the requirements on $m_0$ and $N$ for the static system in terms of a ``hidden symmetry'' of the system: when both $m_0=0$ mod $p$ and  $N+1=0$ mod $p$ are obeyed, the absolute values of the eigenstates are spatially symmetric at sites of the lattice $k$ obeying $k = 0 \mod p$.  When the absolute values of the eigenstates are symmetric at these sites, the first order perturbations to the energies of the system introduced by the imaginary-energy defects vanish, which means that the energy spectrum is real for  $\gamma$ below $\gamma_c$. 

In contrast, our driven model has an unbroken $\mathcal{A}$-symmetric phase for all choices of ${m_0}$ and $N$, as exemplified in Figs.\,\ref{fig:beta1/2} and \ref{fig:criticalvalues}.  As in the static model, the lattice has a phase of unbroken symmetry for $\gamma$ below some critical value $\gamma_c$.   It should be noted that $\gamma_c$ truly represents a phase transition between the $\mathcal{A}$-symmetry broken and unbroken regions: our numerical calculations show that for $\gamma\leq \gamma_c$, the spectrum is real within numerical error and for $\gamma>\gamma_c$ complex quasienergy eigenvalues are present in the spectrum. For $\gamma>\gamma_c$ in the driven model, the entire energy spectrum does not generally become complex all at once: at high frequencies (on the order of $\omega=2\pi$ for the lattices discussed in this work), the highest magnitude quasienergies tend to acquire imaginary parts before the other quasienergies, since the coupling between quasienergy sectors is given by imaginary terms. 
Again, however, a real quasienergy spectrum may be obtained for any choice of $m_0$ as long as $\gamma$ is below the critical value for a given set of parameters.

\begin{figure}
    \begin{subfigure}{0.5\textwidth}
        \centering
        \caption{}
        \vspace{-.05cm}
        \includegraphics[scale=0.55]{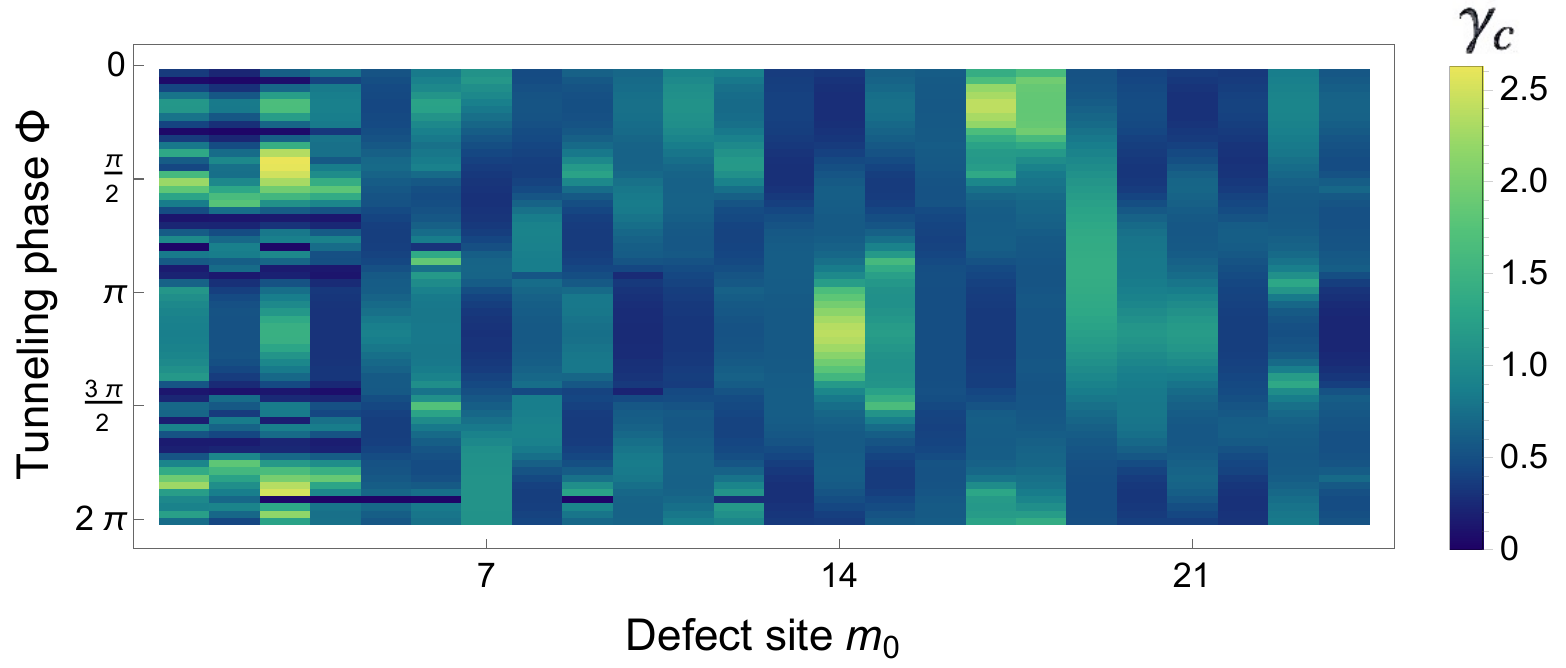}
    \end{subfigure} \newline
    ~ 
    \begin{subfigure}{0.5\textwidth}
        \vspace{-.1cm}
         \caption{}
        \vspace{-.05cm}
        \includegraphics[scale=0.55]{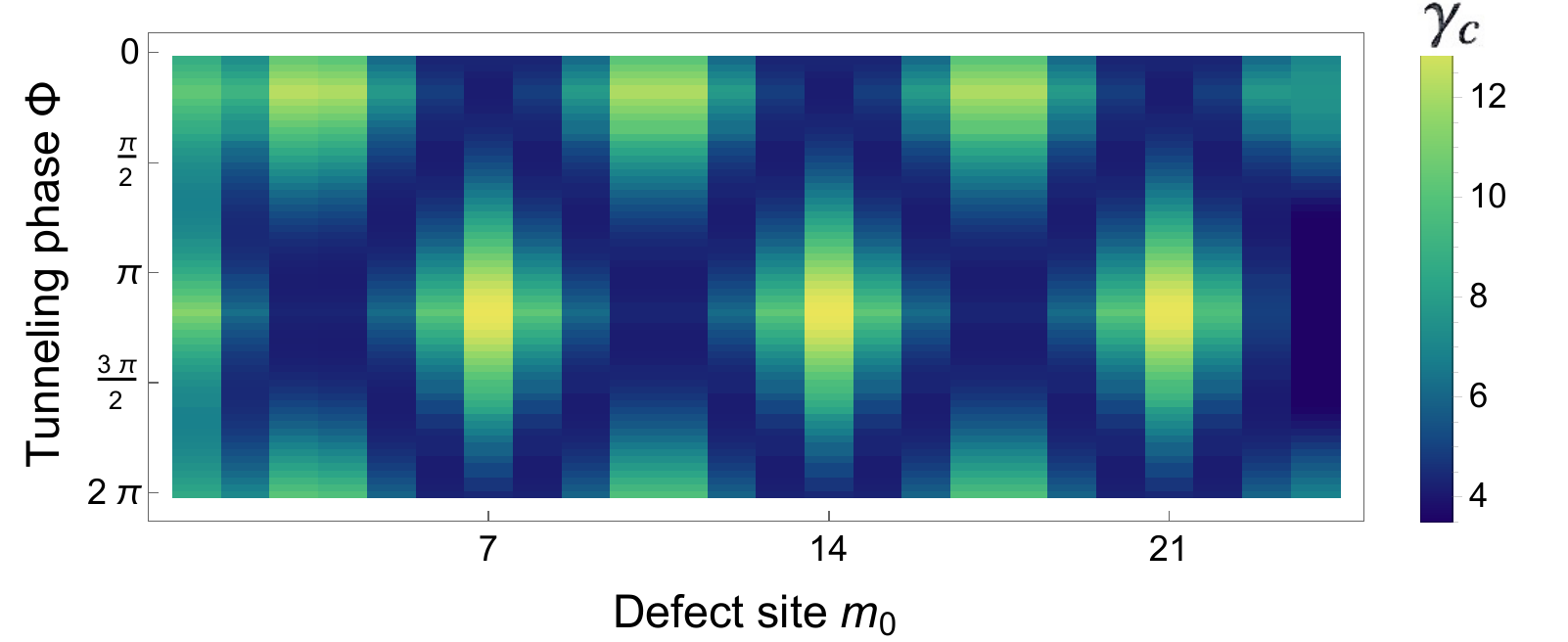}
    \end{subfigure} \newline
        ~ 
    \begin{subfigure}{0.5\textwidth}
        \centering
        \vspace{-.1cm}
         \caption{}
         \vspace{-.05cm}
         \includegraphics[scale=0.56]{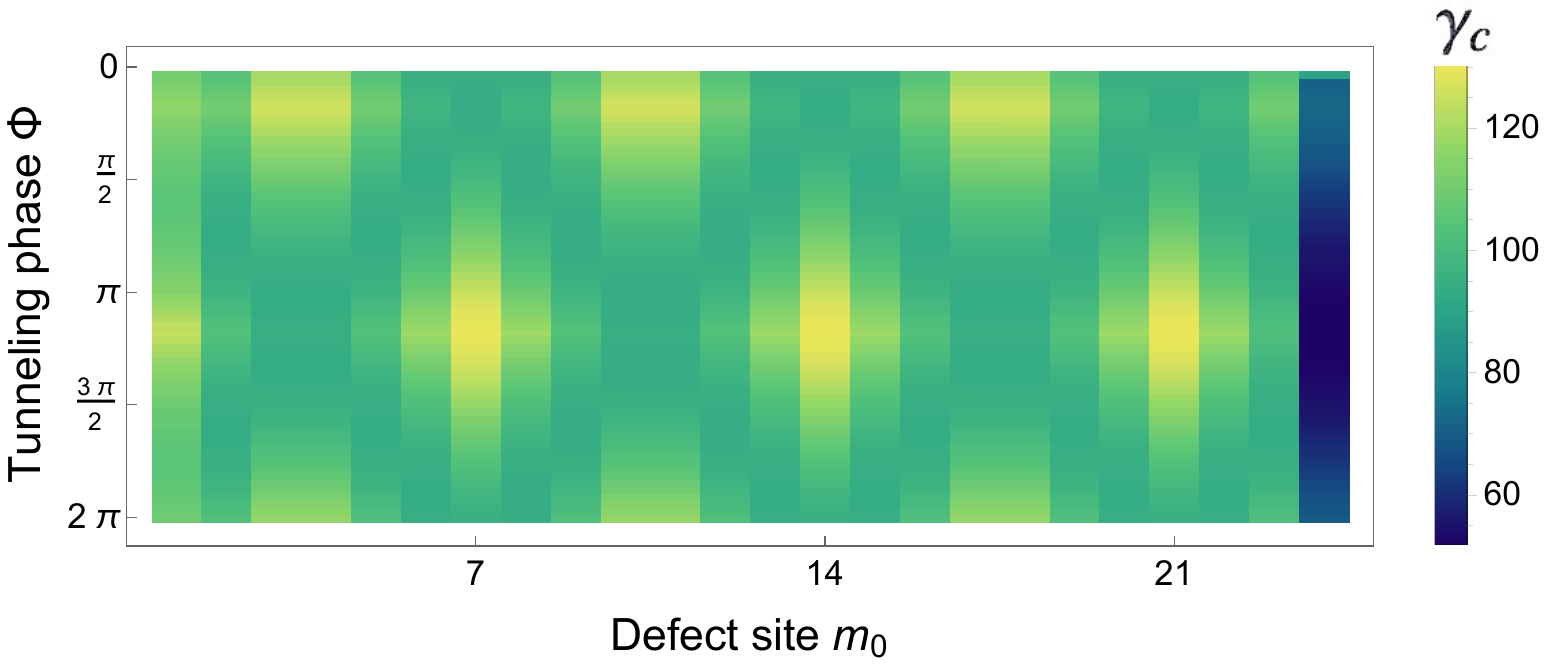}
    \end{subfigure}
    \caption{$\mathcal{A}$ symmetry-breaking threshold $\gamma_c$ for $\beta=1/7$, $N=48$, and $\lambda=0.5$ (a) has $\omega=\pi$, (b) has $\omega=2\pi$, and $\omega=10\pi$}  \label{fig:beta1/7}
\end{figure}

 As shown in Figs.\,\ref{fig:criticalvalues} and \ref{fig:beta1/7}(b)-(c), $\gamma_c$ seems to exhibit a repeating pattern across $m_0$: roughly, when the defects are placed sufficiently far from the edges of the lattice and from each other (i.e. $m_0$ is not too close to 1 or to $N/2$), $\gamma_c$ follows a repeating pattern with the same periodicity of the lattice.  For example, consider two configurations of a lattice with the parameters in Fig.\,\ref{fig:beta1/7}, one with defect location $m_0$ (with the second defect located symmetrically at $\bar{m_0}$) and the other with defect location $m_0'$ (with the second defect located symmetrically at $\bar{m_0}'$).
 If $m_0'=m_0 \mod 7$, Fig.\,\ref{fig:beta1/7} shows that $\gamma_c$ for these to configurations are approximately equal to each other across all $\Phi$.  However, we emphasize that the values of $\gamma_c$ for these two configurations are not exactly equal and but differ by $10^{-4}$ or less. Decreasing $\omega$ causes this correlation to weaken and eventually to disappear, as shown in Fig.\,\ref{fig:beta1/7}.  
 
 This pattern occurs based on the tunneling elements that connect the defects to the rest of the lattice.  
 If the set of four tunneling elements that connect the two defects to the rest of the lattice in the $m_0$ configuration is the same as the set of four tunneling elements that connect the defects to the rest of the lattice in the $m_0'$ configuration, $\gamma_c$ will be approximately equal for these two lattices.  
 This occurs, for example, under the condition mentioned above ($m_0'=m_0 \mod 7$)  and when $m_0=6 \mod 7$ with $m_0'=8 \mod 7$. 
 It should be noted that only the set of four tunneling elements connecting the two defects to the rest of the lattice matters, not which tunneling elements connect which defect. (For example, for the second choice of $m_0$ and $m_0'$ discussed above, the first defect (energy $-i \gamma \cos(\omega t)$) for the $m_0$ configuration will be connected to the lattice by the same tunneling elements as the second defect (energy $-i \gamma \cos(\omega t)$) for the $m_0$ configuration and vice versa.)  This is because complex-conjugating the Floquet Hamiltonian (Eq.\,\ref{eq:Fourier}) reverses the locations of the defects but has no effect on the quasienergies, since the quasienergies come in complex-conjugate pairs as shown in the previous subsection.  As a result,  Fig.\,\ref{fig:beta1/7}(b)-(c) exhibit both a pattern that repeats with the periodicity of the lattice (every 7 sites) and another symmetry about $m_0=0 \mod 7$.  These results hold for all other rational $\beta$ with the same dependence on the tunneling elements.  Why the tunneling elements lead to this pattern is an area for future investigation.  Neither the high frequency expansion nor other perturbative methods (such as Ref.\,\cite{Floquet_perturbation}) lead to a Hamiltonian where, once the conditions on $m_0$ and $m_0'$ are obeyed, the quasienergies of these configurations are exactly the same and differences between them may be treated perturbatively.   
 Again, defect placement near the edges of the lattice does not obey this pattern, and defect placements on the outermost sites give the largest values of $\gamma_c$.  The reason for this is evident from the analysis of the rotating wave approximation shown in the next section. 

\begin{figure}
\includegraphics[scale=.4]{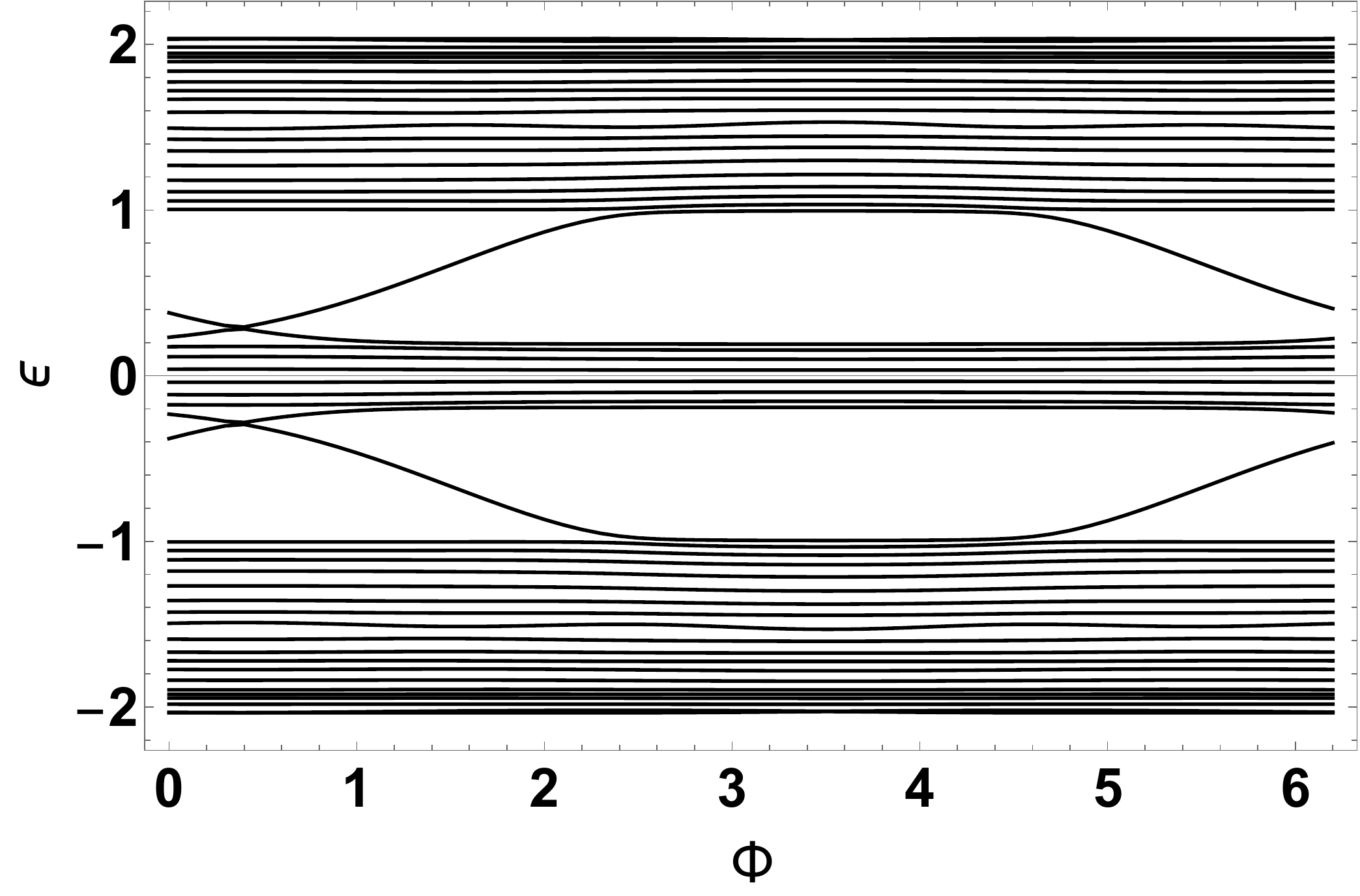}
\caption{Quasienergy spectrum for $\beta=\sqrt{2}$, $N=48$, $\omega=2\pi$, $\lambda=0.4$, $\gamma=3$, and $m_0=12$.  As in the static case, lattices with irrational $\beta$ possess fractional numbers of (quasi)energy bands.} \label{fig:irrational_beta}
\end{figure}

Now, we consider systems where $\beta$ is irrational.  In this case, the tunneling parameters are only quasi-periodic rather than periodic.  In the static analog, the eigenstate symmetry discussed in Ref.\,\cite{Joglekar} is violated and it is therefore impossible to obtain a real energy spectrum.  In contrast, we find that once again the periodically driven system defined in Eq.\,2 possesses a phase of unbroken $\mathcal{A}$ symmetry for all choices of $m_0$ and $N$.   When $\gamma<\gamma_c$, the quasienergy spectra for these configurations are fully real and possess a fractional number of bands. Fig.\,\ref{fig:irrational_beta} provides an example of this.  
Unlike systems with rational $\beta$, though, $\gamma_c$ shows no pattern across $m_0$ because the tunneling elements never repeat. It should also be noted that, as in the time-independent, Hermitian system of Ref.\,\cite{AAH_Hermitian}, the mid-gap modes are localized to the edges of the lattice. 

As discussed above, the quasienergy eigenvalues for both the periodic and the quasiperiodic models depend on driving frequency.  For any $\omega,$ real parts of the first Floquet Brillouin zone quasienergies fall between $-\omega/2$ and $\omega/2$.   Increasing $\omega$  increases $\gamma_c$ for a given set of parameters, but there is no nonzero $\omega$ for which $\gamma_c=0$,  as exemplified in Fig.\,\ref{fig:beta1/7}. So, changing $\omega$ also leads to a phase transition between $\mathcal{A}$ broken and unbroken phases, i.e. the imaginary parts of the spectrum are exponentially suppressed when $\omega$ is higher than some critical value $\omega_c$.  This behavior can be explained in the context of the high frequency approximation shown in Section \ref{sec:series}: at high frequencies the Hamiltonian takes on an effective Hermitian form, and  non-Hermitian higher-order terms carry greater weight as frequency is decreased.  At sufficiently high frequencies, the quasienergies remain constant as $\omega$ is increased further and agree closely with the zeroth order term in the high-frequency expansion shown in Section \ref{sec:series}.  The strengths of imaginary parts of the quasienergy spectrum also depend on $\omega$: in general, once $\omega$ is decreased below $\omega_c$, the magnitudes of the imaginary parts increase as $\omega$ is decreased (or as $\gamma$ is increased).  
%
%
\subsection{\label{sec:series}High frequency expansion}

We have shown that increasing driving frequency $\omega$ can cause a phase transition from a broken to unbroken symmetry phase.  In order to explain this behavior, we use the Floquet-Magnus expansion and the rotating reference frame approximation discussed in Refs.\,\cite{Floquet_more, Floquet_rotating_frame} to show that at sufficiently high frequencies, the lattice of Eq.\,\ref{eq:H} takes on an effective Hermitian form.  

In the laboratory frame, the Floquet operator $U$ is defined by 
\begin{equation}
U= \mathcal{T} e^{-i\int_0^T H(t) dt}, \label{eq:U}
\end{equation} 
where $ \mathcal{T} $ is the time-ordering operator, and obeys $U\ket{\phi_\alpha(0)}=e^{-i\epsilon t} \ket{\phi_\alpha(0)}$.  Transformation to the rotating reference frame is given by the unitary operator $S(t)= \mathcal{T} e^{-i\int_0^t V(t') dt'}$.  Then, the Floquet operator in the rotating frame takes the form $U_r= \mathcal{T} e^{-i\int_0^T H_r(t) dt}$, where $H_r(t)= S(t)^\dagger H_0 S(t)$.  In the high frequency limit, the $H_r(t)$ takes the effective time-independent form 
\begin{equation}
H_{\text{eff}}=H_{\text{eff}}^{(0)}+\frac{1}{\omega} H_{\text{eff}}^{(1)}+\frac{1}{\omega^{2}} H_{\text{eff}}^{(2)}+... \label{eq:series}
\end{equation}
and so $U_r$ is simply $U=e^{-i H_{\text{eff}}T}$ and the quasienergies are given by the eigenvalues of $H_\text{eff}$. \cite{Floquet_rotating_frame}.  To zeroth order, we have $H_{\text{eff}}\approx H_{\text{eff}}^{(0)}=\frac{1}{T}\int_0^T H_r(t) dt$.  For our system, $V(t)$ commutes with itself at different times, and so $H_r(t)$ takes the simple form 
\begin{align}
H_r(t)=\sum_{n} T_{\text{eff}} a_n^\dagger a_{n+1}+h.c.\nonumber \\
\end{align} where \[T_{\text{eff}}= \begin{cases} 
      t_n & n\neq m_0-1,m_0,\bar{m}_0-1,\bar{m}_0 \\
      t_n e^{-\gamma \sin(\omega t)/\omega}& n= m_0-1,m_0\\
      t_n e^{+\gamma \sin(\omega t)/\omega}& n= \bar{m}_0-1,\bar{m}_0.
   \end{cases}\] 
Since $H_r(t)$ is Hermitian, so is its time average $H_\text{eff}^{(0)}$.  Therefore, the quasienergies are real to zeroth order approximation in the high-energy limit.

The precise forms of higher order terms are given in \cite{Floquet_rotating_frame}. First order corrections vanish, and higher-order terms give complex coupling between different sites, which introduce imaginary parts to the energy spectrum.  The fact that finite-frequency corrections take the form of coupling terms agrees with the numerical result from the previous section that defect placements at the edges yield different $\gamma_c$ than other defect placements, i.e. the edges of the lattice couple only to other sites in the lattice only on one side, not both sides.  
We find that at sufficiently high frequencies the zeroth order approximation is in close agreement with the numerically obtained spectrum, and so the spectrum is fully real.  At smaller values of $\omega$, however, the complex higher order terms in Eq.\,\ref{eq:series} do not decrease as rapidly as for higher $\omega$, and so the spectrum becomes complex. Fig.\,\ref{fig:beta1/7} shows how $\gamma_c$ changes with $\omega$.  As stated above increasing $\omega$ increases $\gamma_c$ for a given set of parameters, and $\gamma_c$ is always nonzero.

\subsection{\label{sec:micromotion}Micromotion}

The majority of our analysis concerns quantities that appear as averages over many driving periods.  We now consider the exact behavior of the system over an individual period, referred to as ``micromotion.''  From Ref.\,\cite{Shirley}, a (full, time-dependent) Floquet state is given by
\begin{equation}
    |\psi_{\alpha}(t) \rangle= \sum_q |\phi_{\alpha, q}\rangle \rangle e^{i q \omega t} e^{-i \epsilon_\alpha t}, \label{eq:psi(t)}
\end{equation}
where $n$ enumerates sites, $q$ enumerates photon sectors, and $|\phi_{\alpha,q}\rangle \rangle$ is an eigenvector of the Floquet Hamiltonian in the frequency domain (Eq.\,\ref{eq:Fourier}), and $\epsilon_\alpha$ is the quasienergy eigenvalue corresponding to this eigenvector, all as defined in section \ref{sec:Model and the Floquet formalism}.
%
%

\begin{figure}
    \begin{subfigure}{0.4\textwidth}
    \caption{}
    \includegraphics[scale=0.45]{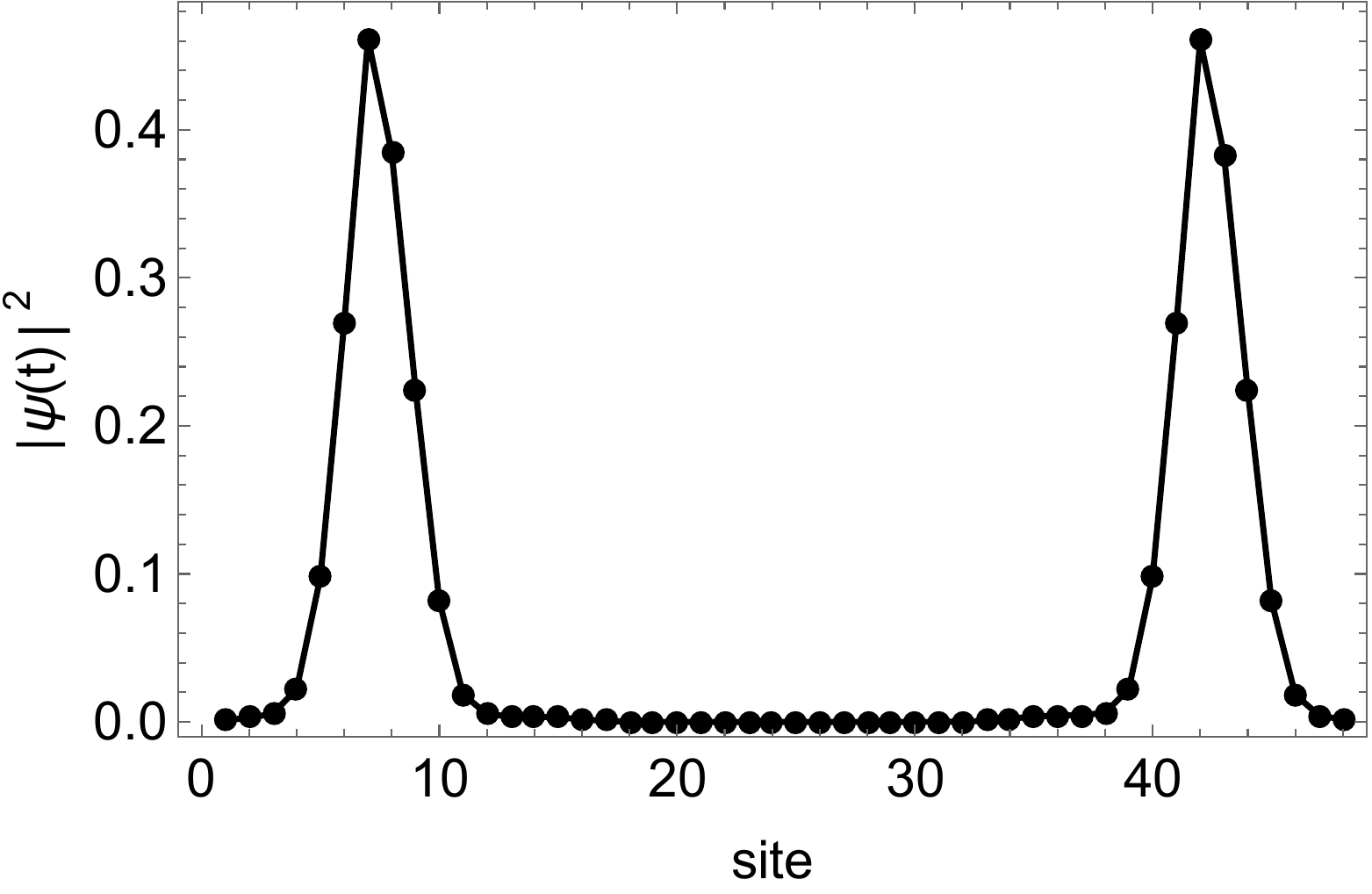}
    \end{subfigure}
    \newline
    \begin{subfigure}{0.4\textwidth}
    \hspace{1cm}
    \caption{}
    \includegraphics[scale=0.43]{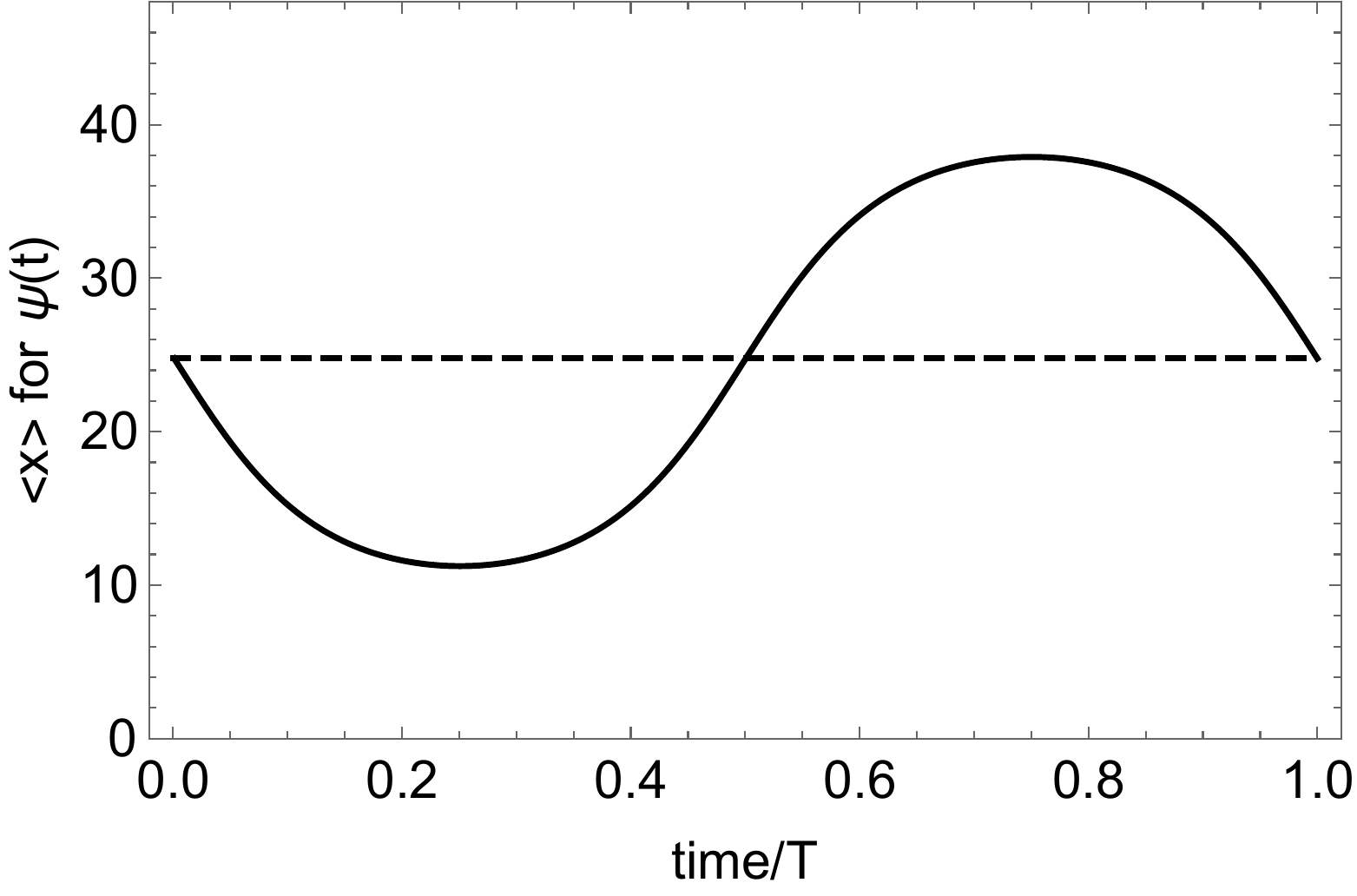}
    \end{subfigure}
    \newline
    \caption{Time evolution of a Floquet state over the course of a period.  (a) shows a time-averaged Floquet state. (b) shows the  expectation value of position for the fully time-dependent Floquet state at each instant of time.   The horizontal dashed line shows the value of $\braket{x}$ for $\psi(t=0)$. (And, since  $\psi(t=T)=\psi(t=0)$, $\braket{x}$ is the same at $t=0$ and $t=T$.) The parameters correspond to those of Fig.\,\ref{fig:beta1/7}b, with  $\beta=1/7$, $N=48$,  $\lambda=0.5$, $m_0=7$, $\Phi=0$, $\gamma=3$, and $\omega=2\pi$ ($T=1$). This state corresponds to the lowest quasienergy in the zeroth photon sector ($\epsilon=-2.57$) and is localized to the two defect sites over the course of the period (though the magnitude of the state at each site changes over the course of a period).}
    \label{fig:Micromotion}
\end{figure}

Given the form of Eq. \,\ref{eq:psi(t)}, each Floquet state remains normalized as long as its corresponding quasienergy is real.  In this case, the factor of $e^{-i \epsilon_\alpha t}$ provides only a phase factor, and so the absolute value of the state is periodic with the same period as the onsite energies.  Similarly, as will be discussed further in Section \ref{sec:topological modes}, $\epsilon=0$ Floquet states remain localized over the entire period.  In fact, we find numerically that their corresponding Floquet modes are localized strongly to the first Floquet Brillouin zone.  (For example, the zero quasienergy, time-averaged Floquet mode shown in Fig.\,\ref{fig:beta1/2} is localized entirely to the first photon sector within numerical precision.  In other words, $|\phi_{\alpha, q} \rangle \rangle$ is nonzero only for $q=0$, and so the full time-dependent Floquet state is static and is equal within numerical precision to the first photon sector of the time-averaged Floquet state shown in Fig.\,\ref{fig:beta1/2}c.)  Other states with real quasienergies, however, may evolve appreciably over the course of a period.  An example of this evolution is shown in Fig.\,\ref{fig:Micromotion}.  Fig.\,\ref{fig:Micromotion}a shows a time-averaged Floquet state, and Fig.\,\ref{fig:Micromotion}b shows, at each instant of time, the expectation value of position for the corresponding fully time-dependent Floquet state. 
As Eq.\,\ref{eq:psi(t)} suggests, Floquet states whose corresponding quasienergy has a nonzero imaginary part do not evolve periodically over time and do not remain normalized; the imaginary part of the quasienergy contributes a factor of exponential growth or decay. 

\section{\label{sec:topological modes}Topological modes in the commensurate model}
%
Having analyzed the reality of the spectrum, we discuss the appearance of topologically protected modes.  As is exemplified in Fig.\,\ref{fig:beta1/2}(a),\,(c), the combinations of parameters that lead to localized zero modes in the time-independent case also lead to localized zero (modulo $\omega$) modes in the time-dependent case.  
The Floquet states corresponding to $\epsilon=0$ mod $\omega$ are localized to the edges of the lattice, as shown in Fig.\,\ref{fig:beta1/2}(c). 
For this combination of parameters and a given $\Phi$ in the topological phase, odd values of $N$ lead to a single zero mode corresponding to a Floquet mode localized at one edge of the lattice.  Even values of $N$ lead to two zero modes when $\Phi<\pi/2$ or when $\Phi> 3\pi/2$ corresponding modes localized at both edges of the lattice.   
These results correspond perfectly to the topologically protected modes in the time-independent systems of Refs.\,\cite{AAH_Hermitian, Yuce}.

We now demonstrate that these localized modes are topologically protected and robust to perturbations. The Floquet Hamiltonian defined in Eqs.\,\ref{eq:H},\ref{eq:H_F} has the particle-hole symmetry 
\begin{equation}
    \mathcal{C} H_F \mathcal{C}^{-1} = -H_F^*, \label{eq:H_Fsymm}
\end{equation}
where the particle-hole operator $\mathcal{C}=\mathcal{D}T$ with $\mathcal{D}$ and $T$ as defined in Eqs.\,\ref{eq:Ddef}-\ref{eq:SDTaction} (The definitions of $\mathcal{C}$ and $\mathcal{A}$ are identical up to a phase shift.)
Particle-hole symmetry guarantees that quasienergies appear in positive-negative pairs.  To show this, we first complex-conjugate the Floquet eigenvalue equation Eq.\,\ref{eq:fake_eval}: 
\begin{align}
    H_F^* \ket{\phi_\alpha(t)}^*= \epsilon_\alpha^* \ket{\phi_\alpha(t)}^*. \nonumber
\end{align}
Acting with $\mathcal{C}$ gives $\mathcal{C}H_F^* \ket{\phi_\alpha(t)}^*= -H_F \mathcal{C} \ket{\phi_\alpha(t)}^*$ and $\mathcal{C}\epsilon_\alpha^* \ket{\phi_\alpha(t)}^*\nonumber =\epsilon_\alpha \mathcal{C}\ket{\phi_\alpha(t)}^*$.  
Therefore, 
\begin{align}
    H_F \mathcal{C} \ket{\phi_\alpha(t)}^* =-\epsilon_\alpha \mathcal{C}\ket{\phi_\alpha(t)}^*
\end{align}
and so for solution to Eq.\,\ref{eq:fake_eval} with quasienergy $\epsilon$, there is a solution with quasienergy $-\epsilon$.  Combined with $\mathcal{A}$ symmetry, this guarantees that in the $\mathcal{A}$-broken phase the quasienergy spectrum has a quartet structure, $(\epsilon,\epsilon^*,-\epsilon,-\epsilon^*)$. 
By the arguments of Ref.\,\cite{Floquet_Majorana_localized_over_time}, the particle-hole symmetry of the Floquet Hamiltonian guarantees the topological protection of the zero energy modes described above and that, furthermore, these modes remain localized over all time.

These modes are, in particular, two-fold degenerate Floquet Majorana modes.  As with the time-independent systems of Refs.\,\cite{AAH_Hermitian, Yuce}, the fermionic operators of Eq.\,\ref{eq:H} can be decomposed into two species of Majorana operators, resulting in a Hamiltonian analogous to the time-independent systems.  In particular, the Hamiltonian takes the form of two Kitaev chains plus two oscillating imaginary-energy terms.  As in the static analog, each Kitaev chain individually supports a Floquet Majorana mode localized at one or both ends in the topological phase \cite{AAH_Hermitian,Yuce}.  Therefore, in the topological phase, the system as a whole possesses a two-fold degenerate Floquet Majorana mode, which acts as a Floquet analog of a Dirac edge mode, and furthermore, the combinations of parameters that lead to doubly degenerate Majorana bound states in the time-independent system also lead to Floquet Majorana modes in this lattice.  (For example, when $\beta=1/2$, this occurs for $|1+ \lambda \text{cos}\Phi|>|1-\lambda \text{cos}\Phi|$. )  Our numerical results confirm this: for example, compare our Fig.\,\ref{fig:beta1/2} to Figs. 2-3 from Ref.\,\cite{Yuce}.  

We can also ascribe to the system a topological invariant by considering the lattice in the absence of defects.  (With the defects, the lattice lacks translational invariance, and so the wavevector $k$ is no longer a good quantum number.)  In this case, a  $\mathbb{Z}_2$ invariant can be calculated for each of the Majorana chains, as discussed further in Refs.\,\cite{AAH_Hermitian,Yuce,Kitaev_2001}, and for appropriate choices of parameters, the $\mathbb{Z}_2$ invariant shows that each Majorana chain is topologically non-trivial and supports a Majorana mode.  The introduction of defects adds a coupling between the two chains, as discussed in Ref.\,\cite{Yuce} and, as we have seen in this work, does not  destroy the topological edge modes.  As in the static analog, the topological modes remain even with the defect because the defect does not introduce coupling between unpaired Majorana operators on the edge of the lattice.
Additionally, because Floquet quasienergies are only defined mod $\omega$, not only are $\epsilon=0$ modes their own particle-hole conjugates, but so are states with $\epsilon=\pm \omega/2$, and therefore Floquet Majorana modes carrying nonzero quasienergy exist in some systems \cite{Floquet_Majorana_proof}.  Our numerical calculations show that such modes do not exist in this model: in order to obtain localized $\epsilon=\pm \omega/2$ modes, the coupling between different frequency sectors of $\mathcal{H}_{n,m}^{p,q}$ in Eq.\,\ref{eq:Fourier} must be sufficiently large in comparison to other energy parameters of the system ($J$ and $\hbar \omega$, where, again, we set $\hbar=1$).  Since this coupling is purely imaginary ($\pm i \gamma$), the spectrum becomes complex, and no  purely real modes with $\epsilon=\pm \omega/2$ exist. 
%

Thus, we have obtained Floquet topological modes in a non-Hermitian, quasi-$PT$-symmetric lattice. 
%
%
\begin{figure*}
    \centering
    \begin{subfigure}{0.5\textwidth}
        \centering
        \caption{}
        \includegraphics[scale=.40]{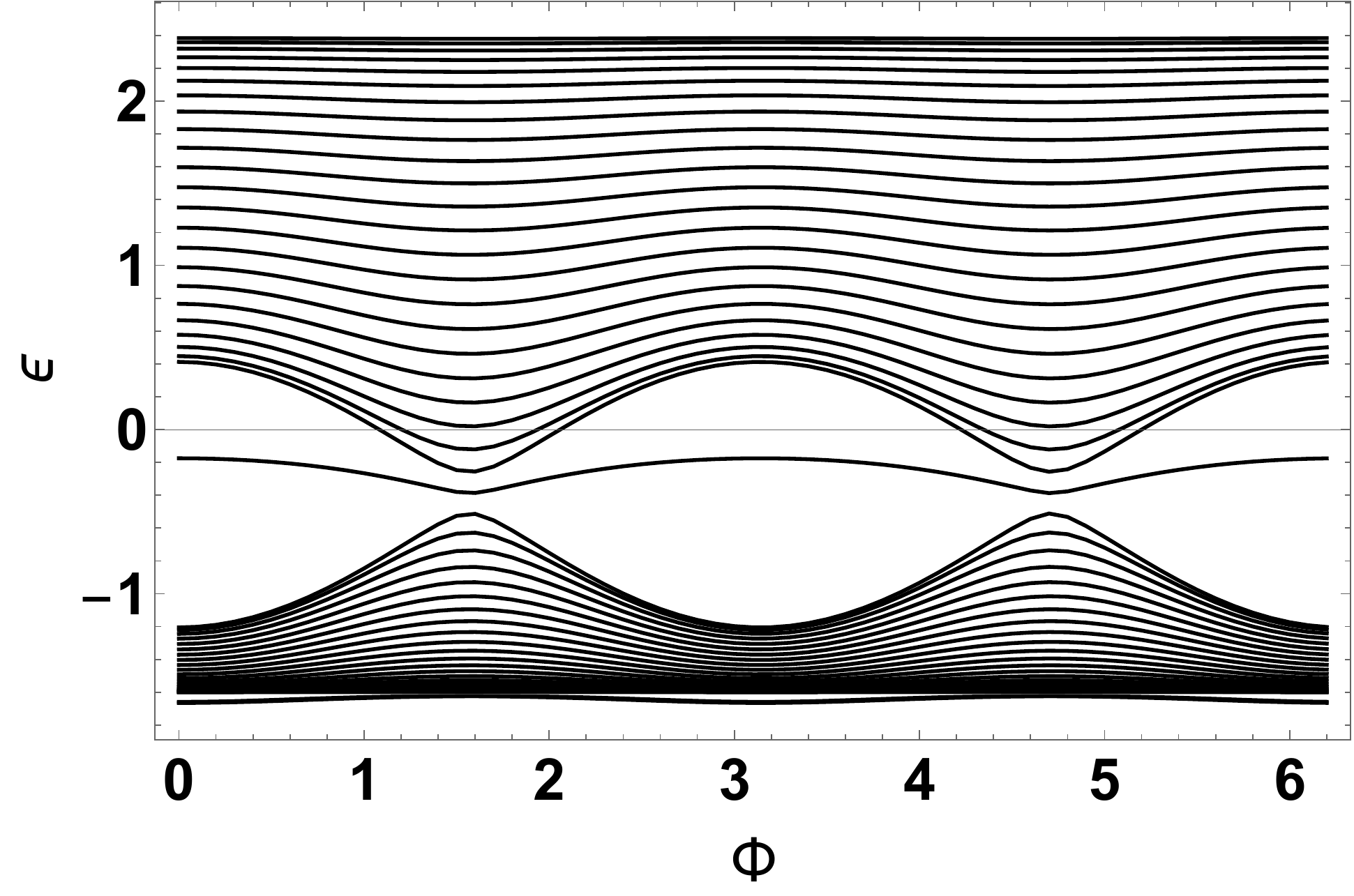}
    \end{subfigure}%
~
    \hspace{.1cm}
    \begin{subfigure}{0.5\textwidth}
        \centering
        \caption{}
        \includegraphics[scale=0.65]{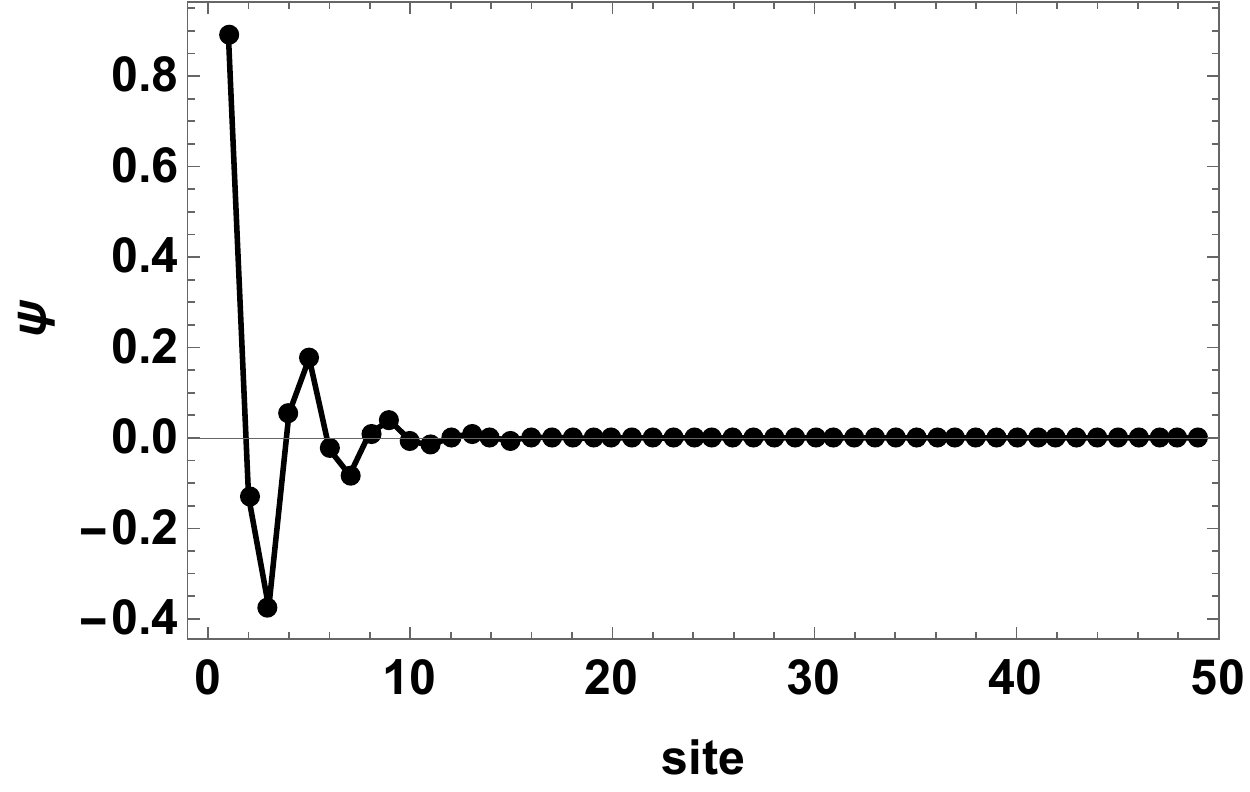}
    \end{subfigure}
\caption{(a) Quasienergy spectrum with next-nearest-neighbor tunneling (Eq.\,\ref{eq:Hnnn}) with $\beta=1/2$, $N=49$, $\omega=2\pi$, $\lambda=0.4$, $\gamma=1$, $m_0=1$, and $t'=.2$.  (b) Time averaged, first Floquet-Brillouin zone $\epsilon=0$ quasienergy state for the spectrum in (a) with $\Phi=0$. Again, this mode is localized to the first Floquet-Brillouin zone, and so only this region is shown.} \label{fig:nnn}
\end{figure*}
We test the topological stability of the zero-quasienergy states by introducing next-nearest-neighbor (nnn) tunneling to the Hamiltonian defined in Eq.\,\ref{eq:H}.  In analogy with Ref.\,\cite{AAH_Hermitian}, we assume the tunneling is site-independent and has strength $t'$. The Hamiltonian of this system is
\begin{align} \label{eq:Hnnn}
H=  -J \sum_{n=1}^{N-1} (1+ \lambda\, \text{cos}(2\pi \beta n+ \Phi)) (a_n^\dagger a_{n+1}+ a_{n+1}^\dagger a_n) \nonumber \\
+ i \gamma \, \text{cos}(\omega t) (a_j^\dagger a_j - a^\dagger_{N-j+1}a_{N-j+1}) \nonumber \\
+ \sum_{n=1}^{N-2} J'(a_n^\dagger a_{n+2} +a_{n+2}^\dagger a_n)
\end{align}
For sufficiently small $t'$, the next-nearest-neighbor (NNN) tunneling term is effectively a perturbation.  
The addition of this term causes the energies of the Majorana modes to be lifted from zero, as shown in Fig.\,\ref{fig:nnn} (a).  However, due to their topological nature, the modes are still localized for sufficiently small perturbations.  For the combination of parameters in this Figure, these modes remain localized until the next-nearest neighbor tunneling parameter is of the same order of magnitude as the nearest-neighbor hopping, $t' \leq 0.6$. 

Because the mid-gap modes remain localized in the presence of local perturbations, they are topologically stable.  This can be understood in analogy with the arguments presented in Ref.\,\cite{AAH_Hermitian}. In particular, the addition of next-nearest-neighbor hopping breaks particle-hole symmetry, and so the mid-gap quasienergies are not guaranteed to be zero precisely.  However, these states can be adiabatically connected to the original zero modes, and so the original localized modes are still present in the presence of next-nearest-neighbor tunneling.

In conclusion, we have presented a periodically varying, non-Hermitian generalization of the off-diagonal AAH model.  In the topological phase, zero mod $\omega$ quasienergy modes appear and are protected by particle-hole symmetry, making this the first generalized $PT$-symmetric system to exhibit a Floquet topological phase.  We have also shown how the definitions of $PT$ symmetry and generalizations of $PT$ symmetry can be applied within the Floquet formalism, and that, like in the static case, the preservation of these symmetries can replace the requirement of Hermiticity.   For our system, transitions between the symmetry broken and unbroken phases can be achieved by changing the driving frequencies, as well as the usual parameters found in static analogs.  We hope that this work will shed light on the intersection between non-Hermiticity and Floquet topological phases and will allow for further investigation in this novel area of study.  
\section{\label{sec:Acknowledgements}Acknowledgements}
E.N.B. thanks Noah Graham, Chris Herdman, Taylor Hughes, and Kohei Kawabata for enlightening discussions.

\bibliographystyle{apsrev4-1}
\bibliography{Bib_test}

\end{document}